\documentclass[preprint2]{aastex}

\newcommand{\degree}{^{\circ}}

\slugcomment{Version Submitted to APJ}

\shorttitle{A symmetric inner cavity in the HD 141569A transitional disk}
\shortauthors{J. Mazoyer et al.}

\begin{document}

\title{A symmetric inner cavity in the HD~141569A circumstellar disk}

\author{J.~Mazoyer\altaffilmark{1,2} and A.~Boccaletti\altaffilmark{2} and \'E.~Choquet\altaffilmark{1} and M.~D.~Perrin\altaffilmark{1} and L.~Pueyo\altaffilmark{1} and J.-C.~Augereau\altaffilmark{4,5} and A.-M.~Lagrange\altaffilmark{4,5} and J.~Debes\altaffilmark{1} and S.~G.~Wolff\altaffilmark{3}}
\altaffiltext{1}{Space Telescope Science Institute, 3700 San Martin Dr, Baltimore MD 21218, USA}
\altaffiltext{2}{LESIA, Observatoire de Paris, CNRS, UPMC and Univ. Paris Diderot, 5 place Jules Janssen, 92190 Meudon, France}
\altaffiltext{3}{Johns Hopkins University, 3400 North Charles St., Baltimore, MD, USA}
\altaffiltext{4}{Univ. Grenoble Alpes, Institut de Plan\'e{}tologie et d\'{}Astrophysique (IPAG) F-38000 Grenoble, France}
\altaffiltext{5}{CNRS, Institut de Plan\'e{}tologie et d'Astrophysique (IPAG) F-38000 Grenoble, France}
\email{jmazoyer@stsci.edu}

\begin{abstract}
Some circumstellar disks, called transitional or hybrid disks,
present characteristics of both protoplanetary disks (significant amount of gas) and debris disks (evolved structures around young main-sequence stars, composed of second generation dust, from collisions between planetesimals). 
Therefore, they are an ideal astrophysical laboratory to witness the last stages of planet formation.
The circumstellar disk around HD~141569A was intensively observed and resolved in the past from space but also from the ground but the recent implementation of high contrast imaging systems opens new opportunities to re-analyze this object.
We analyzed Gemini archival data from the Near-Infrared Coronagraphic Imager (NICI) obtained in 2011 in the H band, using several angular differential imaging techniques (classical ADI, LOCI, KLIP). These images reveal the complex structures of this disk with an unprecedented resolution. We also include archival Hubble Space Telescope (HST) images as an independent dataset to confirm these findings.
Using an analysis of the inner edge of the disk, we show that the inner disk is almost axisymmetrical. The measurement of an offset towards the East observed by previous authors is likely due to the fact that the Eastern part of this disk is wider and more complex in substructure. Our precise re-analysis of the eastern side shows several structures including a splitting of the disk and a small finger detached from the inner edge on the southeast. Finally, we find that the arc at 250 AU is unlikely to be a spiral, at least not at the inclination derived from the first ring, but instead could be interpreted as a third belt at a different inclination.
If the very symmetrical inner disk edge is carved by a companion, the data presented here put additional constraints on its position. 
The observed very complex structures will be confirmed by the new generation of coronagraphic instrument (GPI, SPHERE). However, the full understanding of this system will require gas observations at millimetric wavelengths.
\end{abstract}


\section{Introduction}
\label{sec:intro}

The majority of stars around which exoplanets have been directly imaged also harbor dust disks \citep{Lagrange2009, Marois2008, Rameau2013}. Interestingly, these young planetary systems ($<100$ Myr) feature structures in the dust distribution, such as gaps, warps, or cavities, often recognized as signposts of planets \citep{Ozernoy2000}. For a long time, the observation of circumstellar disks has been the privilege of the Hubble Space Telescope (HST). However, the improvement of instrumental techniques (coronagraphs with dedicated adaptive optics systems) as well as powerful post-processing methods have recently allowed taking advantage of the full capabilities of 8-m class telescopes for the analysis of these objects. HST data have also benefited from such data processing developments \citep{Soummer2011,Soummer2014}. 

HD~141569A is a B9.5V star (V = 7.12, H = 6.86) and is one of the closest Herbig Ae stars \citep[116 pc, ][]{vanLeeuwen2007}. This young star \citep[$5 \pm 3$ Myr, ][]{Weinberger2000} is possibly orbited by two low-mass stellar companions at $\sim 1000$ AU \citep{Weinberger2000,Reche2009}. A circumstellar disk, classified as transitional \citep{Fisher2000}, has been  imaged with several instruments at various wavelengths. First it was observed simultaneously in 1999 with HST/NICMOS in the F160W (1.4 -- 1.8 $\mu$m) filter \citep{Augereau1999_hd141} and F110W (0.8 -- 1.4 $\mu$m) filter \citep{Weinberger1999}. A few years later, this analysis was completed by other HST instruments: with HST/STIS in the visible (0.2 -- 1.1 $\mu$m) by \cite{Mouillet2001} and with HST/ACS in visible wavelengths (three filters between 0.3 and 1.0 $\mu$m) by \citep{Clampin2003}. Finally, it was observed from the ground, with the 200-inch Palomar telescope in K band \citep{Boccaletti2003}, and more recently, once augmented with an extreme AO system, at H and K band \citep{Wahl2013}.

These observations revealed several prominent structures. The disk, seen with an inclination of $\sim55 \degree$, is composed of two ring-like belts at $\sim 210$ AU and $\sim 380$ AU from the star, separated by a relatively darker gap in which lies an arc in the East direction, attributed to a potential spiral arm \citep{Clampin2003}. The distances in AU in this paper were measured using the recent 116 pc distance, according to \cite{vanLeeuwen2007} (and corrected for papers anterior to this publication).
Several authors \citep{Clampin2003,Boccaletti2003} reported an offset toward the East of the two belts of 20-30 AU with respect to the star. Two symmetrical large-scale spirals, possibly triggered by the stellar companions\citep{Augereau_papaloizeau2004, Ardila2005} and/or an outer planet \citep{Wyatt2005,Reche2009} start at the edge of the outer ring. In scattered light, the surface brightness (SB) of the disk rapidly drops inside the inner belt, suggesting a region depleted in dust. However, IR photometry with IRAS also suggests the presence of dust at regions $<100$ AU \citep{Augereau1999_hd141}, which is confirmed by resolved observations in the mid IR \citep{Fisher2000,Goto2006,Moerchen10, Thi2014}. Finally, the important presence of cold and hot CO gas has been confirmed multiple times (e.g. \citealt{Thi2014} and Pericault et al., submitted).

The Near-Infrared Coronagraphic Imager \citep[NICI, ][]{Toomey2003} has already been successfully used to detect disks around HD\,142527 \citep{Casassus2013}, HD\,100546 \citep{Boccaletti2013b} and HD\,15115 \citep{Mazoyer2014}. Recently, \cite{Biller2015} reported observations of HD\,141569A disk system, using archival data from 2011, from which they concluded that the disk belts are 
likely inclined differently ($51 \degree$ or $45 \degree$ for the inner ring, $47 \degree$ for the outer ring) with respect to the line of sight. In addition, they measured a small but still significant offset of the inner disk ($4 \pm 2$ AU). Here, we present an independent and alternative analysis of the same NICI data as well as HST/NICMOS reprocessed data, which reaches a different set of conclusions as we describe below. 

Observations and data processing are described in Section~\ref{sec:obs}. In Section~\ref{sec:morph}, we first present a morphological analysis of the disk, then conduct a simplified forward modeling study of the inner part of the disk in Section~\ref{sec:forward}. Finally, we summarize and discuss  possible interpretation in Section~\ref{sec:ccl}.

\begin{figure*}
\begin{center}
 \includegraphics[width = \textwidth]{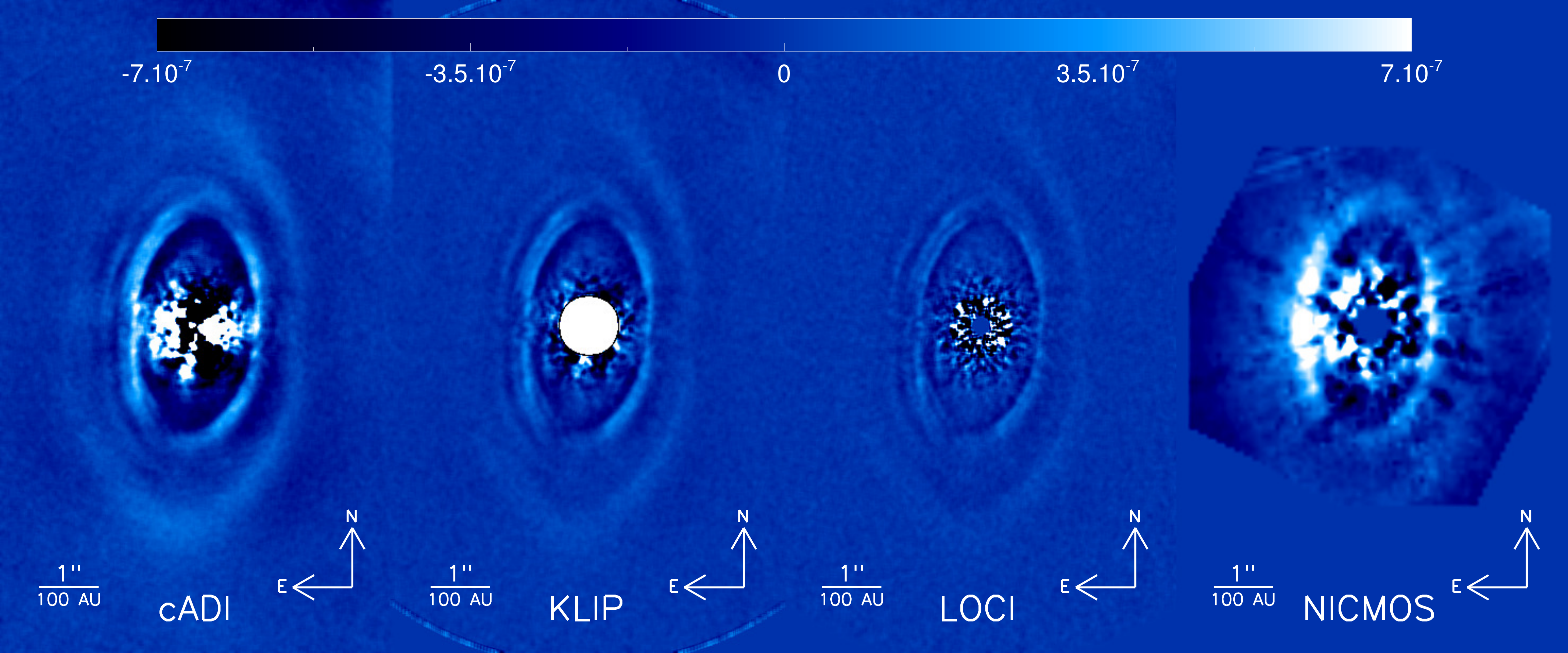}
 \end{center}
\caption[ADI_disk]
{\label{fig:ADI_LOCI_KLIP_disk}From left to right: NICI images processed with cADI, KLIP and LOCI, and NICMOS image re-processed with KLIP, as discussed in the text. The images are smoothed by the PSF width in each case to remove noise. For NICI images, the color scales indicate the contrast with respect to the maximum intensity in the unsaturated image of the star through the semi-transparent mask. For the NICMOS image, we used the precise photometric calibration of this instrument to produce a contrast to the star.}
\end{figure*}

\section{Observations and data reduction}
\label{sec:obs}

\subsection{Gemini/NICI data}
\label{sec:nici_obs}

We used the archival data from NICI, the near-IR (1-5 $\mu$m) AO-assisted dual-band camera installed  at the Gemini South telescope from 2009 to 2013. Several semi-transparent Lyot coronagraph masks of various radii (0.22" to 0.90") are available. The pixel size of this instrument is 17.9 mas. 
Different filters can be inserted in the two channels to allow spectral differential imaging \citep{Racine1999} of brown dwarfs and young exoplanets, but in the scope of this paper, we only use angular differential imaging reduction \citep[ADI, ][]{Marois2006}, taking advantage of the numerous images taken in the pupil tracking mode of the instrument.

HD~141569A was observed several times during NICI's operational lifetime in 2009, 2010 and 2011. We reduced all the data available in the archive but retained the latest observations since they have better quality in terms of AO-correction and achieved contrast. These data were obtained on 2011 May 3 as part of the GS-2011A-Q-501 program, using the filters CH4S ($\lambda = 1.652$ $\mu$m and $\Delta \lambda = 0.066$ $\mu$m) and CH4L ($\lambda = 1.578$ $\mu$m and $\Delta \lambda = 0.062$ $\mu$m) in the two channels. The radius of the coronagraph used in this case is of 0.22''. The number of frames acquired (124) and the total variation of the parallactic angle during the observations (59.6$\degree$) are large and explain the unprecedented quality of the images.

The data reduction follows the prescription provided in \cite{Boccaletti2013a} and \cite{Mazoyer2014}. This includes dark subtraction,  flat fielding, and bad pixels correction. The frame registration makes use of the star's attenuated image behind the semi-transparent mask. A few images were removed from the sequence because they were not adequately centered onto the coronagraphic mask. We ended up with 121 frames of 2 coadds of 30.02 s each for a total integration time of 7264.84 s. We obtained the photometric calibration from the attenuated star image assuming a transmission of $5.94 \pm 0.02$ mag in H band \citep{Wahhaj2011}. The Full Width at Half Maximum (FWHM) of the Point Spread Function (PSF) is 4.5 pixels ($= 85$ mas $\simeq 1.5 AU$). 

The temporal data cube is processed with a series of ADI based algorithms which affect differently the morphology of the disk structures. The self-subtraction is relatively large for such a disk as it is extended and close to face-on \citep{Milli2012}. Therefore the parameters of the 3 treatments we used were carefully selected for different purposes.

Classical ADI \citep[cADI, ][]{Marois2006} limits the self subtraction and preserves the disk photometry. Indeed, Figure~\ref{fig:ADI_LOCI_KLIP_disk} (left) shows that the outer disk is well retrieved in this case. 

In addition, we used Karhunen-Lo{\`e}ve Image Projection algorithm \citep[KLIP, ][]{Soummer2012} in the range 25 to 320 pixels ($\simeq 0.45-5.70$''), excluding the central region dominated by coronagraphic leakage around the mask. For the reasons given above, the reference coronagraphic image to be subtracted is made of three Karhunen-Lo{\`e}ve eigen-vectors (KL = 3) of the principal component basis (of size 121 as the number of images in the initial cube) in order to avoid a large subtraction of the disk itself. The KLIP algorithm is well suited for forward modeling \citep{Soummer2012}, following the method already used in \cite{Mazoyer2014}. 

Finally, we used the Locally Optimized Combination of Images \citep[LOCI, ][]{Lafreniere2007} algorithm already proved efficient for extended objects in many prior cases \citep{Thalmann2010, Lagrange2012, Rodigas2012, Currie2012, Boccaletti2013b, Mazoyer2014}. However, in contrast to the ``conservative'' approach proposed by \cite{Thalmann2010} to avoid the self-substraction of the disk, we used it with a more aggressive set of parameters to retrieve local structures. Using the notation of \cite{Lafreniere2007}, we took a small optimization area $N_A = 300$ PSF footprints, set the radial width of the subtraction zone $d_r$ to $5 \times$ the FWHM, and the minimum radial motion of the image of a point source between two consecutive selected frames was selected as $N_\delta = 0.75 \times$ FWHM. The ratio of the radial to azimuthal width was set to 1. These parameters determine the inner radius of the corrective zone, and the outer radius was set to 300 pixels ($\simeq 5.3$''). With this reduction, the outer ring is almost completely self-substracted, but small structures appears in the inner disks that will be described in the next sections.

As we are not interested in spectral differential imaging (devoted to giant planets where methane is present), the final products of ADI based algorithms applied to the two narrow band data sets were summed to improve the signal-to-noise ratio (SNR). The resulting images for these three algorithms (cADI, KLIP, LOCI) are displayed in Figure~\ref{fig:ADI_LOCI_KLIP_disk}.

\begin{figure*}
\begin{center}
 \includegraphics[height=6cm]{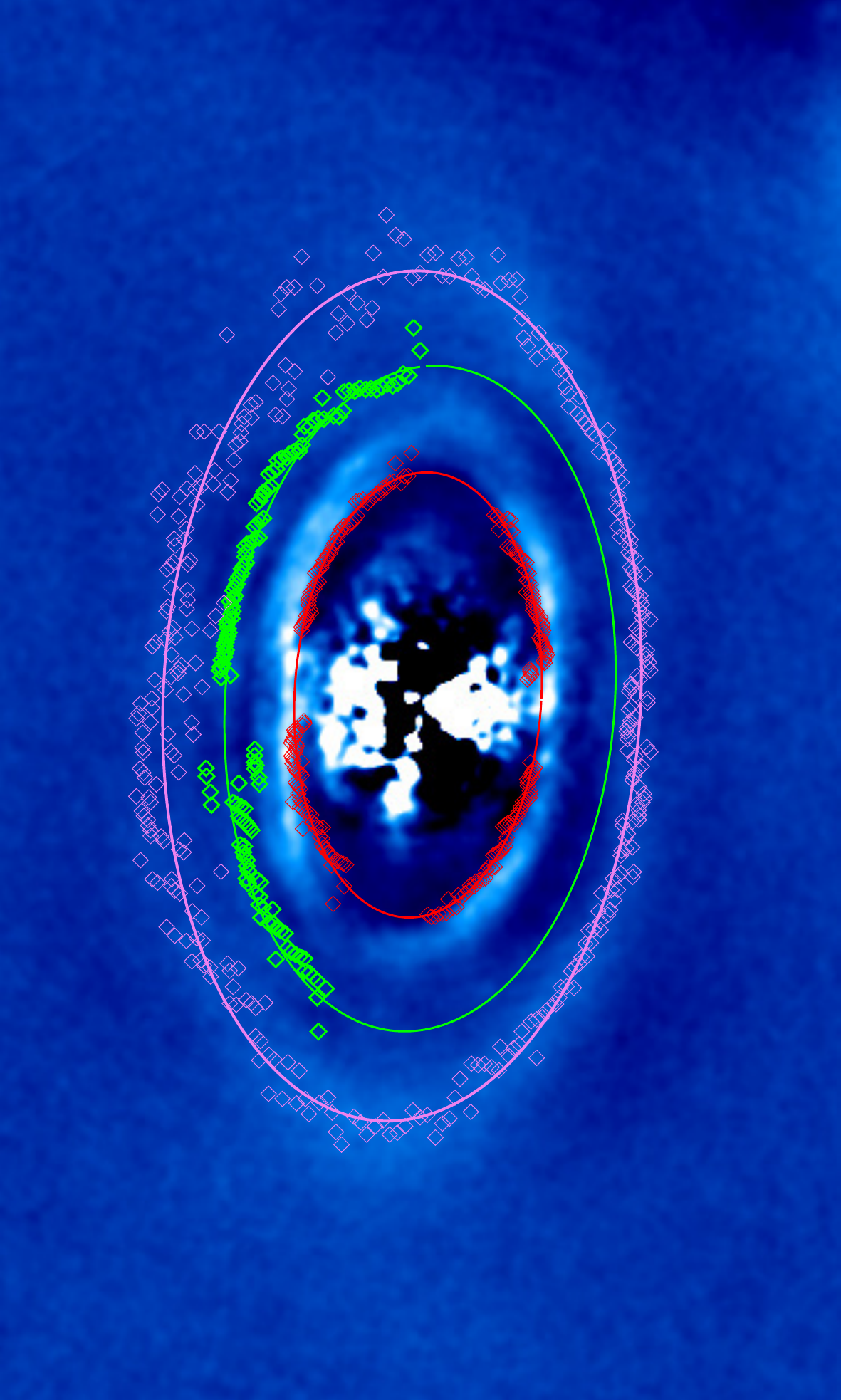}
 \includegraphics[height=6cm]{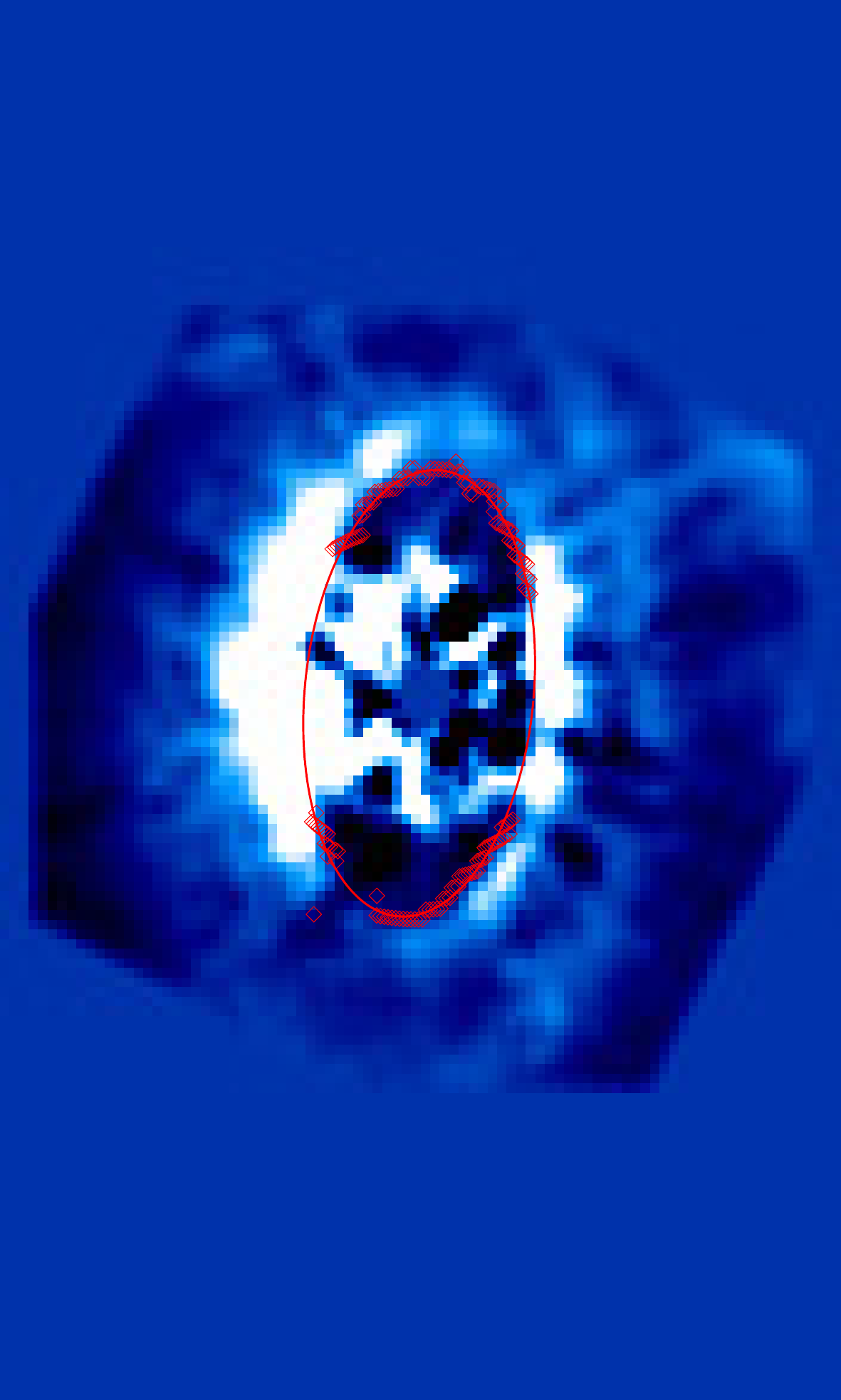}
  \includegraphics[height=6cm]{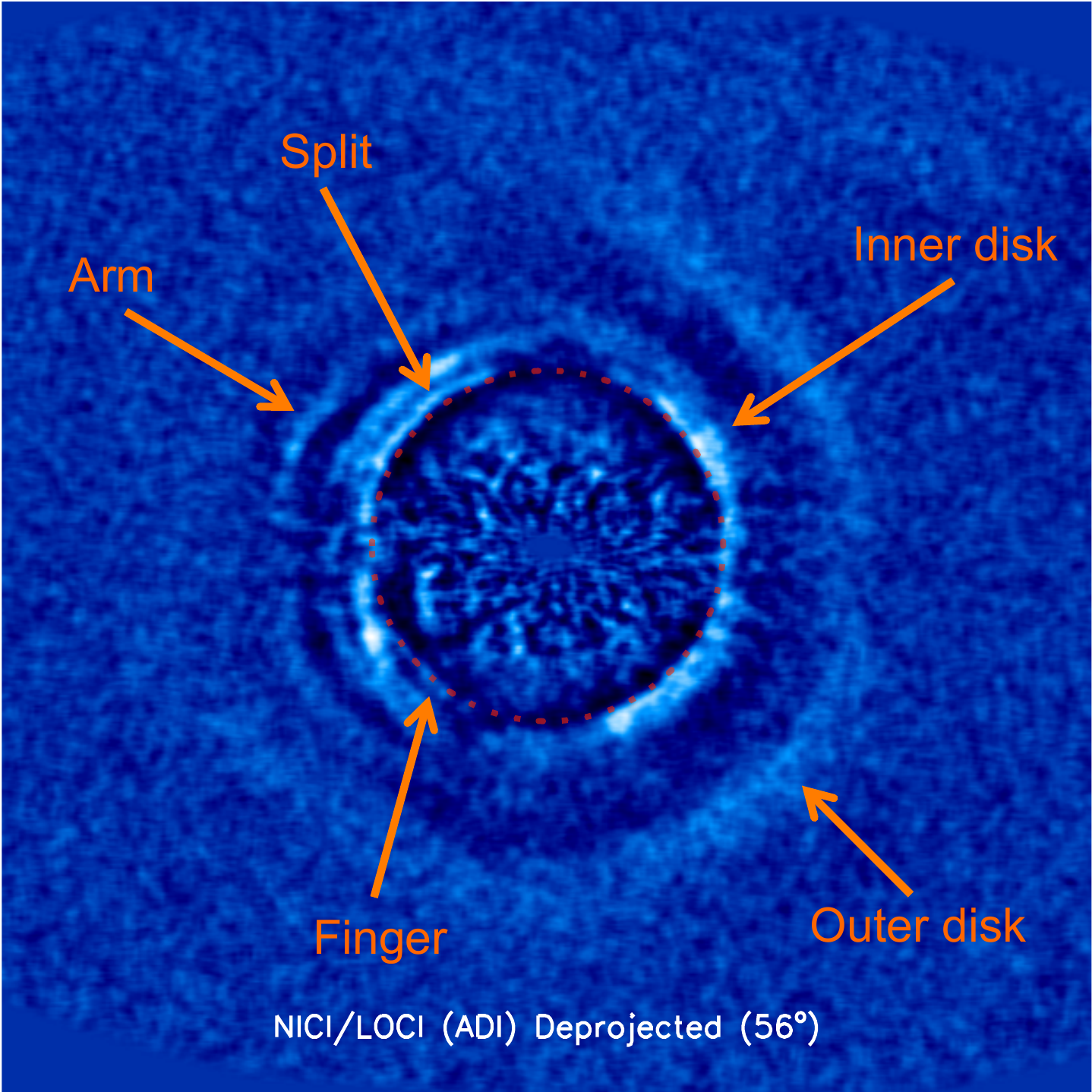}
 \end{center}
\caption[ADI_disk]
{\label{fig:ellipsefit}Left: NICI/cADI image, on which we fitted the position and ellipse fitting of 1) the inner edge of the cavity (in red), 2) the arm on the East (in green), 3) the outer disk (in mauve). Center: NICMOS/KLIP (RDI) image, on which we fitted the position and ellipse fitting of the inner edge of the cavity (in red). Right: Annotated deprojected ($56\degree$) image of the NICI/LOCI (ADI) image.}
\end{figure*}
\subsection{HST/NICMOS data}
\label{sec:nicmos_obs}

HD~141569 was observed on 1998 August 17 using the coronagraphic mode of HST NICMOS, as part of a program aimed at imaging circumstellar disks around four pre-main sequence stars close to the main-sequence stage (HST program GO-7857, PI A.-M. Lagrange). The coronagraph is in the NIC2 channel and includes a 0.3\arcsec-radius focal plane stop in a Lyot coronagraph. The detector sampling in the NIC2 channel is 75.7 mas per pixel, fully Nyquist-sampled beyond 1.7~$\mu$m. The dataset on HD~141569 includes 6 exposures of 144~s and was obtained in a single visit to the star, with the detector oriented at $-117\degr$ from North. The images were acquired with the F160W filter (pivot wavelength $\lambda_p=1.603~\mu$m, full-width at half-maximum of the bandwidth $FWHM=0.401~\mu$m), which is similar to the ground-based $H$ band but slightly broader; it includes the bandpass of the NICI dataset described above. These data were originally analyzed and published by \cite{Augereau1999_hd141}, who presented the first image of the outer ring of HD~141569, simultaneously with---and independently from---\citet{Weinberger1999} who presented NICMOS coronagraphic observations in the F110W filter (close to the J band).

The F160W dataset was processed as part of the \emph{Archival Legacy Investigation of Circumstellar Environments} (ALICE) project (HST program AR-12652, PI R. Soummer), a comprehensive re-analysis of the NICMOS coronagraphic archive using advanced post-processing methods. The data were previously re-calibrated with contemporary flat-field frames and observed dark frames and with a better bad-pixel correction as part of the archival \emph{Legacy Archive PSF Library And Circumstellar Environments} (LAPLACE) program \citep[HST program AR-11279, PI G. Schneider, see][]{Schneider2010}. 
The data were then processed with the KLIP algorithm, by building PSF libraries from reference star images only, using the Reference star Differential Imaging (RDI) processing strategy. We used the ALICE reduction pipeline \citep{Choquet2014} to assemble a library of 276 images from 46 reference stars, register the images of HD~141569 and of the references, then subtract from the science images synthetic PSFs generated with the KLIP algorithm \citep{Soummer2012}. The subtracted PSFs were built out of the 39 first principal components of the library (KL = 39), as a compromise between speckle noise suppression and throughput maximization on the disk.  The final combined image results from the average of all 6 reduced images. 
We scaled the image using the precise calibration of the NICMOS detector which allow the conversion of pixel counts in magnitude. To compare with the NICI image, we then divided by the magnitude of the star in the F160W filter. The resulting image is presented in the right panel of Figure~\ref{fig:ADI_LOCI_KLIP_disk}. The reanalysis yields tremendous improvement in the NICMOS image: \cite{Augereau1999_hd141} only imaged the outer disk, while we completely retrieve the inner disk. Note that the ALICE processing is optimized for the very close environment of the target stars and does not include the full field of view of NICMOS, hence the truncation of the image shown. 

The reanalysis of the 1999 F110W images did not present the same improvement and they were not included in this paper. 

\section{Morphological analysis}
\label{sec:morph}

\subsection{Image analysis}
\label{sec:images}

The NICI images as well as the reprocessed NICMOS image allow detection of both the inner and outer belts at all position angles, while some parts where masked in previous observations due to imperfect subtraction of the stellar residuals. Our reduction also achieves sharper views of fine details in the disk than prior reductions. The previously discovered structures, in particular the suspected spiral inside the disk, can now be discussed from this new perspective. 

In all images of Fig.~\ref{fig:ADI_LOCI_KLIP_disk} we identify the following structures starting from the outer part and going inwards:
\begin{itemize}
\item[a.] The outer belt, which appears most clearly in the cADI image. The East side is fainter and more diffuse than the West side, but the self-subtraction of the disk tends to attenuate this difference with respect to previous observations hence reinforcing the intensity along the major axis, while the minor axis is pinched as expected from ADI artifacts \citep{Milli2012}. In the northeast, the outer belt can be seen to extend into an arc-like feature. 
The two large scale and wide opened spirals identified by \citet{Clampin2003} and possibly triggered by the flyby of the stellar companions are barely detected at the North and South of the outer ring edges. Since these have very low signal to noise ratio we do not discuss these patterns further. 

\item[b.] A clearly defined arc-like pattern is present in between the two main rings to the East, already observed by \citet{Mouillet2001} and \citet{Clampin2003}. Interestingly, it possibly echoes the other arc seen, at low S/N, in the outer ring in the same direction. The intermediate arc seems to originate from the northern part of the inner ring and develops to the southern part. The fact that this region was partly obscured in previous HST data could be the reason why it was rather considered as a spiral. Indeed, although we can see it all along the East part of the disk, it doesn't look attached to the southern part of the inner ring.

\item[c.] The inner ring is recovered at all position angles and reveals a complex morphology. The most important characteristic is that the East side is clearly split into two narrow rings/arcs as opposed to the West side which appears as a single thinner arc. This splitting can be clearly detected from PA = $10\degree$  to PA = $160\degree$. In retrospect, this feature can also be seen in the STIS and ACS images, but was either unnoticed or not commented upon by either \citet{Mouillet2001}, \citet{Clampin2003} or \cite{Biller2015}. A small finger feature in the southeast of the inner edge is detached from the inner ring, between PA = $130\degree$  to PA = $160\degree$. These two features (splitting, finger) are not described in \cite{Milli2012} and are not symmetrical with respect to the major axis. Therefore, they are very unlikely to be due to the ADI processing. However, the effect of ADI on such thin and small structures has not been investigated yet. Finally, the inner disk also presents differences in luminosity. The major axis ansae are actually dimmer than the rest of the disk, which cannot be explained only by ADI treatment (see Section~\ref{sec:forward}).

\item[d.]The inner edge of the inner ring is very steep (see Section~\ref{sec:forward}). Although such an intensity gradient can be produced by ADI processing, the RDI image from NICMOS independently corroborates the sharpness of this edge. The inner cavity appears depleted in dust in the NICI and NICMOS images, and we do not detect the inner dust previously identified using the mid and far-IR observations. However, the stellar diffraction residuals limit the detection of any real features inside $~1"$ for NICI and $1.2''$ for NICMOS, which is larger than the expected position of this dust ($<$100 AU).

\end{itemize}
\begin{figure*}
\begin{center}
 \includegraphics[height=6cm, trim= 1.3cm 0.4cm 0.7cm 0.3cm, clip = true]{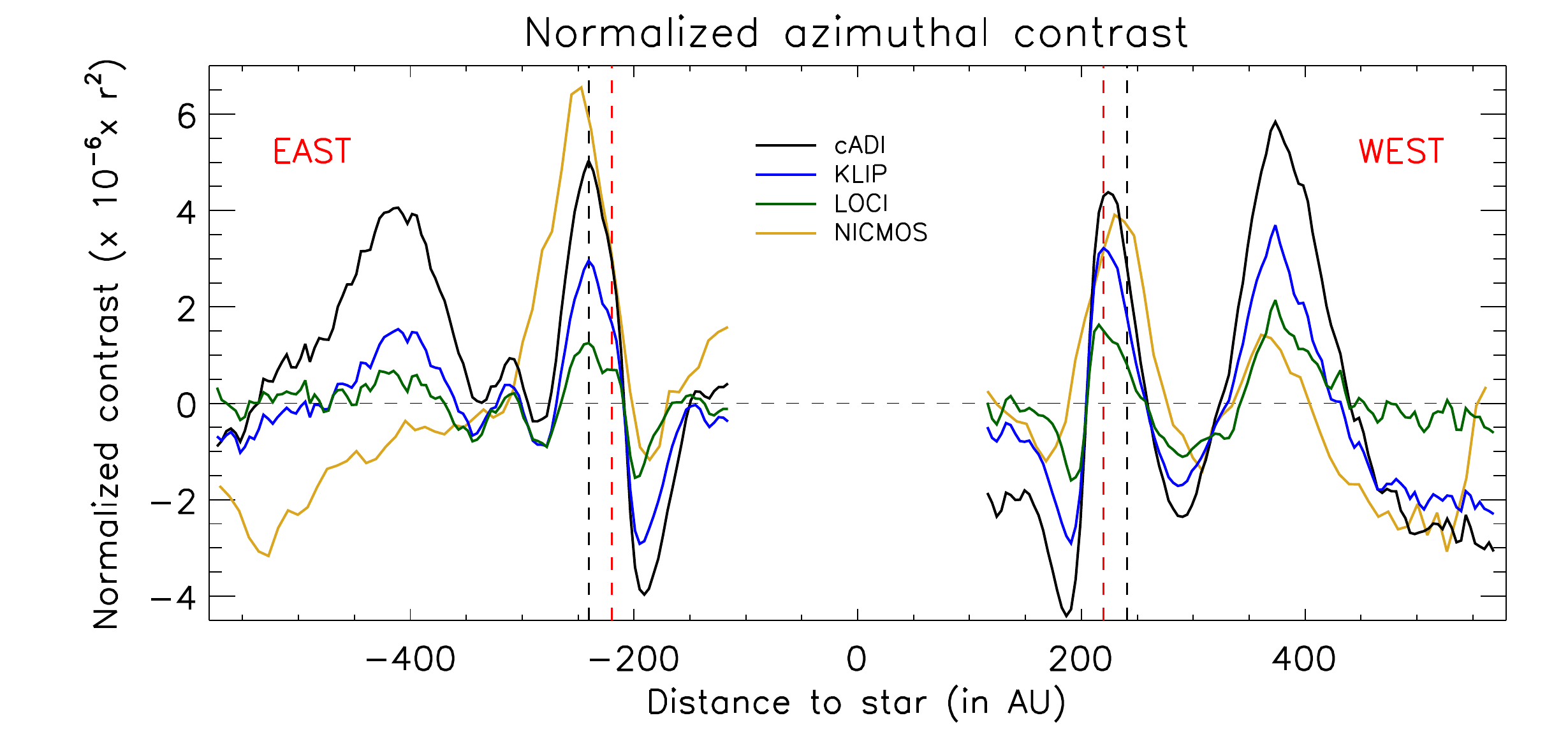}
 \end{center}
\caption[ADI_disk]
{\label{fig:normalized_contrast} Mean azimuthal contrast on the deprojected disk, normalized by the square of the distance to the star. The two vertical dashed line represents the positions of the two part of the eastern split disk.}
\end{figure*}
One general aspect of the disk morphology is that the East side appears definitely more complex than the West side. Hence the registration of the rings, which are wider on the East than on the West, may impact the determination of the disk inclination and location with respect to the central star.  
To avoid any confusion, we choose to measure the disk inclination from the edge of the inner ring, which we measure as the locations where the derivative of the intensity is the maximum, at every angle. This provides a diagnostic that is less biased by disk substructure (arcs) than using the position of maximum brightness along a ring. We excluded the locations that are strongly affected by the ADI processing, namely the minor and major axes \citep{Milli2012}. These points are shown in Fig.~\ref{fig:ellipsefit} (left) in red color. The corresponding locations of the inner edge are then fitted by an ellipse, considering the PA, the inclination and the position of the center as free parameters. The best-fitted ellipse, also represented in red on this figure show an inclination of $i = 56.4\degree$ $\pm 0.5$ and the PA of $356.3\degree$ $\pm 0.6$ are consistent with previous measurements with HST images, but differ significantly from the ones obtained by  \cite{Biller2015} on the same data. We show this discrepancy in Appendix A and demonstrate that our parameters better fit the inner disk. The error bars are derived from the dispersion obtained with the different algorithms. However, contrarily to most previous publications, the ellipse presents very little offset: $25\pm 4$ mas towards East and $20\pm 7$ mas towards North. These offsets are less than 1.5 pixels and less than half the resolution element. They are ten times smaller than the ones found by \cite{Boccaletti2003}. Our assumption is that the previously un-resolved structures in the East part of the inner belt presented in paragraph .c of this Section resulted in a slightly larger East part compared to West part and biased the estimation. Our analysis fitting the maximum of the derivative at every angle instead of the maximum of the intensity removes this bias.

We repeated the same analysis on the NICMOS data (Fig.~\ref{fig:ellipsefit}, center). The parameters obtained are slightly different ($i = 59 \degree  \pm 1$ and PA = $354\degree \pm 0.5 $). They also show very little offset of the inner disk ($37 \pm 11$ mas towards the East and $9 \pm 7$ mas towards the North). The error bars were obtained using several reductions with a different number of KL of the same disk. The small offset is confirmed. However, due to the fact that the NICMOS data have lower resolution and more stellar residuals, we use the values found in the NICI data instead.

We also plot in green on the East part of Figure~\ref{fig:ellipsefit} (left) the position of the intermediate arc, measured by fitting radially a Gaussian function. Although this arc appears to be attached on the northern ansae and not on the South, we find that it is unlikely to be a spiral, at least at the same inclination plane. Using the inclination measured previously, we calculate for every angle the 'derotated radius', meaning the radius of this arc for every angle, assuming it is in the same plane. The variation of the radius $r$ as a function of $\theta$ is symmetrical with respect to the $\theta = $ PA $ + 90 \degree$, which is not consistent with a linear $r = \alpha \theta$ spiral function. We fitted an ellipse centered on the star on this arc and obtain an inclination of $54\degree$, which is coherent with a belt in a slightly different plane. However, as we have only information on the East side, we cannot rule out a belt in the same plane but offset.

Finally in Fig.~\ref{fig:ellipsefit}, we indicate the position of the outer disk in purple. We only use the cADI data in this analysis because the outer disk is mostly self-subtracted with the two other algorithms. The East part of this disk is faint, however, we can drawn an ellipse on these data using the same PA and inclination as for the inner ring in the NICI image and with a offset towards the East (0.15'') with large uncertainties. 

\subsection{Disk de-projection}
\label{sec:deproj}

We deprojected the disk using the derived geometry ($i = 56\degree$, PA =$356\degree$). The results are shown in Figure~\ref{fig:ellipsefit} (right). The dotted red line represents a centered circle of radius 205 AU. On the same figure, we annotated the different components of the disk. This confirms that the inner edge of this disk appears circular and non-offset with this deprojection. 

To highlight the structure of the disk, we then plotted the mean azimuthal contrast of the deprojected  disk, multiplied by the square of the distance to the star to compensate for the inverse square drop in illumination by HD~141569A. The major axis ansae were excluded to limit the influence of ADI treatments. We also excluded the closest regions to the star because of the high speckle noise at these location. The resulting intensity profiles as a function of the distance to the star (in AU) are presented in Fig~\ref{fig:normalized_contrast}, for the three algorithms. 

The vertical red lines, located at $r_1 = 213$ AU (assuming a star at 116 pc) corresponds to the first component of the eastern inner ring (closer to the star), which appears symmetrical around the star. The contrast obtained in the two directions differs by 25\% to 40\%, depending on the method of reduction. The black line, at $r_2 = 234$ AU, correspond to the second peak on the East Part of the disk, which have no equivalent on the West part. This feature is particularly obvious in the LOCI treatment. We see that a lower angular resolution (NICMOS in yellow) tends to blur these features into a wider arc, which explains the difficulties of previous authors in deriving the offset and inclination of the inner disk from peak intensity with HST data. 

Also on the East part, we can spot the intermediate arm's position at $r_3 = 305$ AU. No significant features appears at this location on the West.  The position of the outer disk is seen peaking at $r_4 = 363$ AU in the East part, and $r_5 = 417$ AU on the West part. The difference between these positions gives an offset of $(r_5 - r_4)/2= 25$ AU towards the East of the outer ring, consistent with previous authors. The outer ring also presents an important brightness asymmetry (50\% to 70\%, depending on the reduction method). Note that the effects of anisotropic scattering is inclination-dependent \citep{Mulders2013}, which makes the brightness asymmetry in the deprojected image difficult to interpret. 

\section{Forward modeling}
\label{sec:forward}

\begin{figure}
\begin{center}
 \includegraphics[width=0.52 \textwidth ]{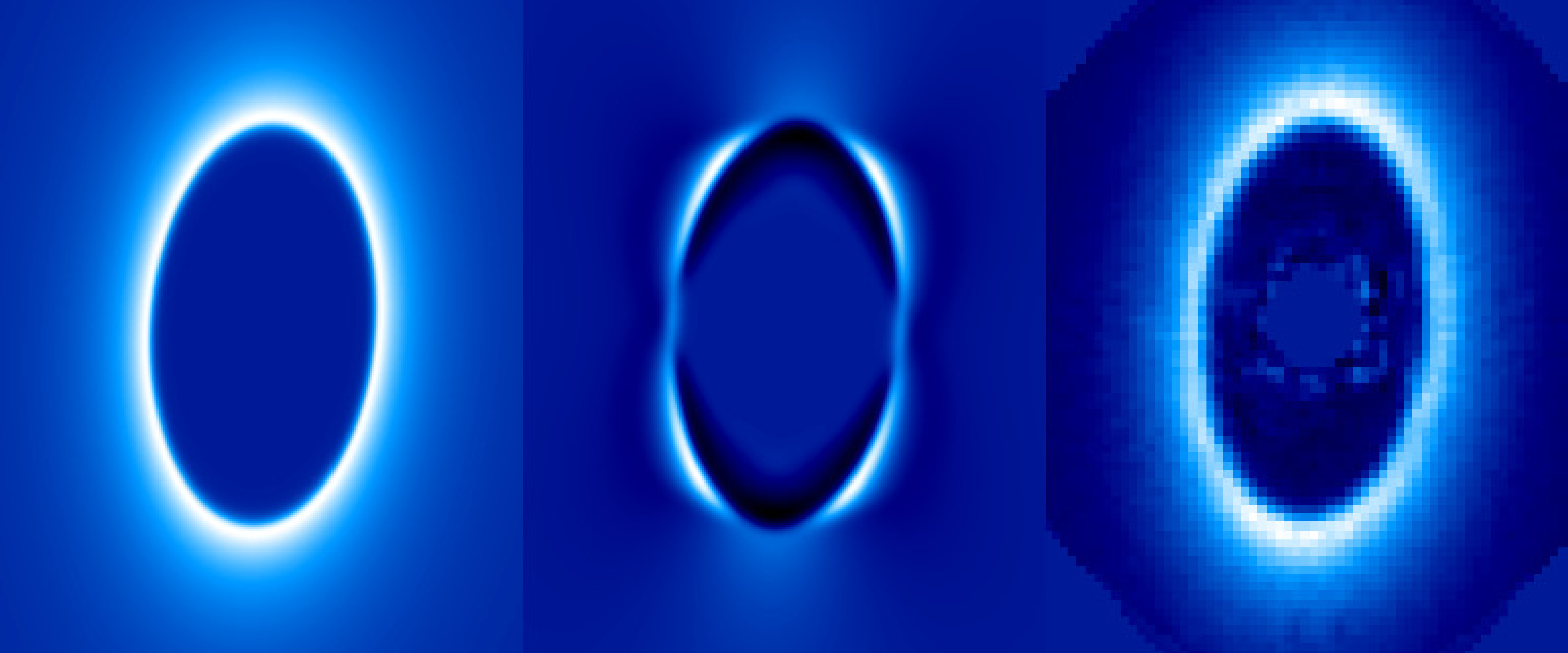}
 \includegraphics[width=0.42 \textwidth ]{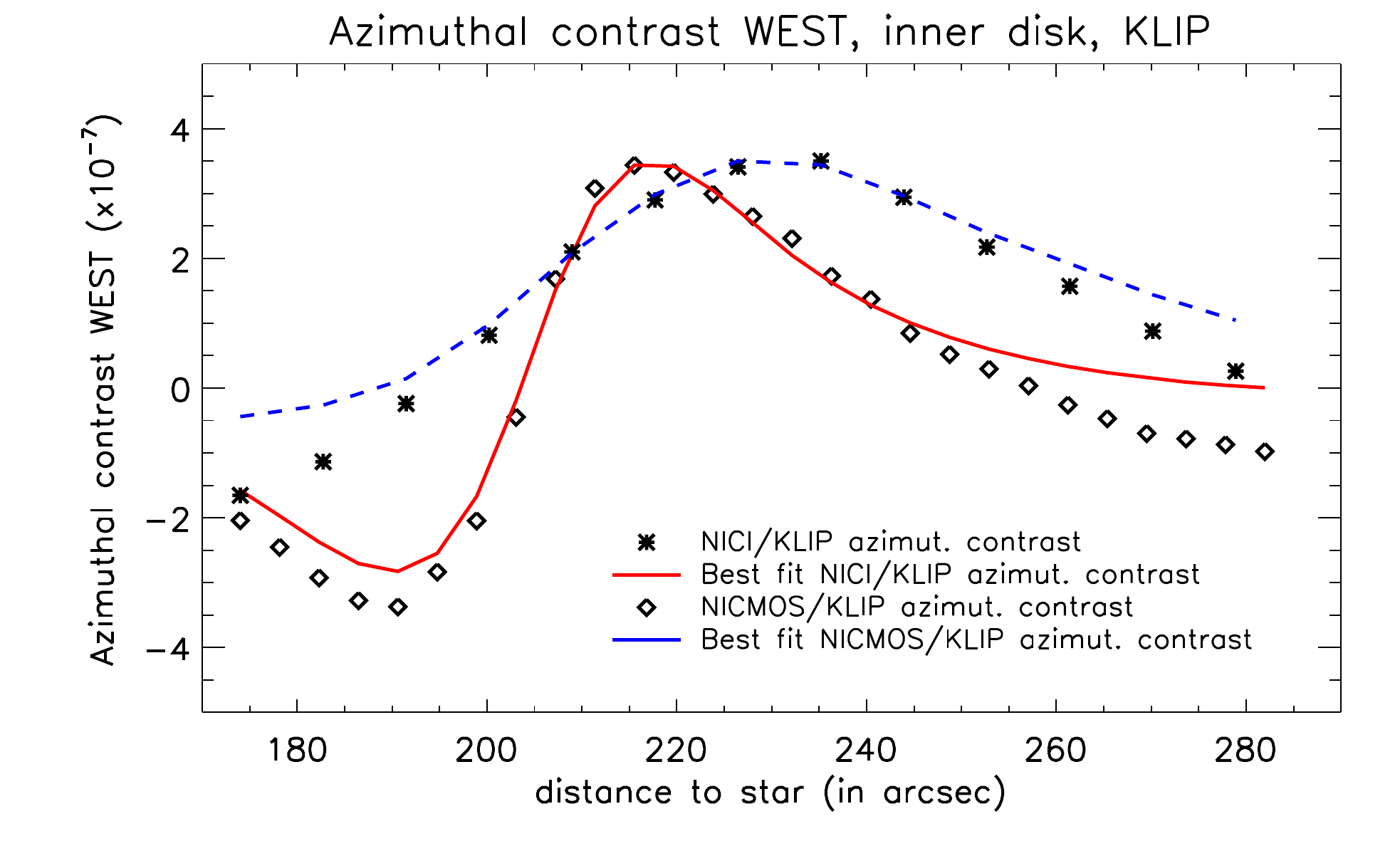}
 \end{center}
\caption[ADI_disk]
{\label{fig:forward_model} Left: best fit model. Center, left: Forwarded version of this model through NICI PSF convolution and KLIP/ADI treatment. Center, right: At the same scale, forwarded version of this model through NICMOS PSF convolution and KLIP/RDI treatment. Comparing these images with the Figure~\ref{fig:ADI_LOCI_KLIP_disk}, especially the ansae, we can rule out if an effect is only due to ADI or not. Right: azimuthal contrast on the West part and best fit model.}
\end{figure}

Given the complexity of the morphology (split of the inner ring, arm in the East part, offset of the outer ring), a full modeling approach is beyond the scope of this paper. Instead we focus on the inner ring and more precisely on the West side of the image, which appears as a simple belt in our data. The goal being to determine the exact position and steepness of the inner edge of this belt. 

We used the GRaTer model \citep{Augereau1999_hr47_grater,Lebreton2013} to simulate disks. The simplified version of this code is essentially geometrical and allows us to produce scattered light images of optically thin radially symmetric disks. We introduce these simulated disks in our data with the forward-modelling technique described in \cite{Soummer2012} and already used in our previous papers \citep{Boccaletti2013a, Mazoyer2014}. We first determine the Karhunen-Lo{\`e}ve vectors and eigenvalues by processing the HD~141569A cube for each filters. We then produce a grid of models using the GRaTer code, convolved with the PSF and then decomposed on this basis using the same eigenvalues, for each filter separately. We finally take the median in each filter to obtain the simulated disk after post processing. There are 6 parameters in the GRaTer model: the belt is located at a reference position $r_0$ from the star, the slopes inward and outward of this belt determined by radial power laws with indexes $\alpha_{in}$ and $\alpha_{out}$. Finally, disk geometry is also given by the PA, the inclination $i$, and the opening angle $h = H/R$. We also assumed isotropic scattering. Using the morphological analysis presented in Section~\ref{sec:morph}, we fixed the first two parameters to PA=$356\degree$, $i=56\degree$. In accordance with a previous study of the internal part of the disk \citep{Thi2014}, we assume a flat disk and set $h$ to $1\%$. We simulated a grid of models varying the last three parameters (and the global intensity of the disk) and used a $\chi^2$ minimization using the azimuthal contrast on the West part of the inner ring as a cost function to obtain the best-fit model (Figure~\ref{fig:forward_model}, left). 

In Figure~\ref{fig:forward_model} (center, left), we present the forward-modeled version of the best fit model through PSF convolution and KLIP/ADI treatment. We show how it fits the azimuthal contrast on the West part in Figure~\ref{fig:forward_model} (right). The outer slope is $\alpha_{out} = -2$, the inner slope is extremely steep as expected, with $\alpha_{in} = 37$ and the ring peaks at $r_0 = 207$ AU. The steepness of the inner slope makes it difficult to retrieve and the precise value is maybe not very meaningful, as it appears to be steeper than the resolution element, even with the good resolution of NICI. We can set a lower limit of $\alpha_{in} > 20$ for which the $\chi^2$ function doubles compared to its minimum value. Using the same argument, we can draw $r_0 = 207 \pm3$ AU. The outer slope value is possibly influenced by the effect of the ADI on the outer disk (not modelled here) and is probably a lower limit. A more precise determination of this outer slope would require a more complete modeling of the full disk, which is out of the scope of this paper.

Although it has strong stellar residuals, we also used the NICMOS image to do forward modelling, which gave a similar result:  outer  slope  of $\alpha_{out} = -7$,  inner slope of $\alpha_{in} = 39$ ($\alpha_{in} > 19$) and $r_0= 208 \pm 7$ AU. The forwarded version of this model through PSF convolution and KLIP/RDI treatment is presented in Figure~\ref{fig:forward_model} (center, right). As expected the RDI process is much better preserving of disk structure and brightness than the ADI process. 

From the comparison of the two forward modelled versions of the same model and with the real data in Figure~\ref{fig:ADI_LOCI_KLIP_disk}, we can analyze whether a feature is due only to the ADI or not. For example, the pinching in the minor axis ansae 
of the inner and outer rings in NICI data does not appear in the NICMOS. A quick look on the Figure~\ref{fig:forward_model} show that this is indeed an effect of the ADI, as noted by previous authors \citep{Milli2012}. On the contrary, the ansae, particularly in the south, are dimmer than the rest of the belt. While ADI can affect the photometry of ansae in a ring-like disk \citep{Milli2012}, in this case, the azimuthal intensity drop is both detected at near IR with ADI (Figure~\ref{fig:forward_model}, center left) and in the visible with RDI (Figure~\ref{fig:forward_model}, center right) which make it a reliable feature. A decrease in disk-scattered light at this location also appears in the ACS image \citep{Clampin2003}. Surface density variations along the ring or anisotropic scattering could be responsible for this feature.

The steep slope of the inner ring can be the result of a planetary mass companion carving the disk and clearing the inner part. \cite{Wisdom1980} obtained a relation between the mass ratio of the planet and star ($M_p/M_s$) and the position of the planet with respect to the star ($a$) and to the inner edge of the disk ($\delta a$): $\delta a / a = 1.3(M_p/M_s)^{2/7}$. Assuming the position of the inner edge $\delta a + a$ is between 201 and 208 AU, we obtained important constrains in the distance-mass relation.

The limits of detection were already derived by \cite{Biller2015} on the very same data. In the separation range 160-208 AU, they found no planet more massive than $M_p<1.4 M_J$ \citep[COND model, ][]{Baraffe2003} or $M_p<3.4 M_J$ \citep[DUSTY model, ][]{Baraffe2002}. Therefore, we can determine the corresponding separation $\delta a$ between an undetected planet and the edge of the cavity compliant with these measurements. If a planet is responsible for sculpting the inner edge it should be orbiting between 179 AU from the star and the disk inner edge in the COND model and 173 AU from the star and the disk inner edge in the DUSTY model.

\section{Summary and discussion}
\label{sec:ccl}

\begin{figure}
\begin{center}
 \includegraphics[height=4cm]{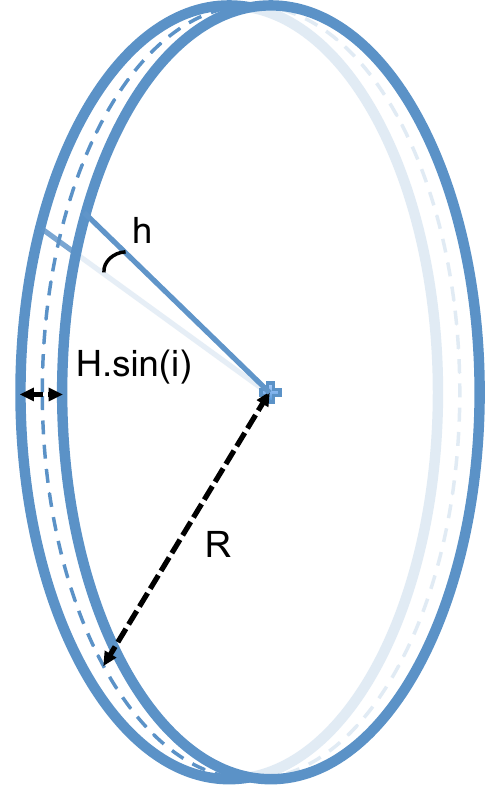}
 \end{center}
\caption[ADI_disk]
{\label{fig:schema_thick} Schematic of one hypothesis for explaining the apparent splitting of the ring on the East side, assuming an optically thick ring with nonzero thickness $h$. However, analysis from \citep{Thi2014} showed that this interpretation is unlikely.}
\end{figure}

\begin{table}[]
\centering
\begin{tabular}{|c|c|c|c|c|c|c|}
\hline
i($\degree$) & PA ($\degree$)& offset W (mas) & offset N (mas) & $r_0$ (AU) &  $\alpha_{in}$ & $\alpha_{out}$ \\
\hline
$56.4 \pm 0.5$ & $356.3 \pm 0.6$ & $25\pm 4$ & $20\pm 7$ &  $207 \pm3$ & 37 ( $> 20$) & $<-7$ \\
\hline
\end{tabular}
\caption{Best fit parameters resulting of the analysis of the inner ring of HD~141569A}
\label{tab:my_label}
\end{table}

We analysed the inner disk orbiting HD~141569A and found an inclination of $i = 56\degree$ by studying the position of the inner edge. This is the main point in which our results differ from those obtained by \cite{Biller2015} using the same data. They used a simplified model of the disk (two simple rings) and performed their fitting by a minimization on the whole image. However, the complexity of the additional substructure on the East part (split inner disk, arc) led to smaller inclination ($i = 51\degree$). Here we show that although the very complex structures in the East part of the inner disk make it difficult to fit by a centered ellipse, one can find an inclination for which the inner edge of this ring is almost perfectly circular and with very little offset relative to the central star.

We then used a geometrical model to constrain the West part of the inner disk. In this simplify case, we therefore avoid the East part of the disk, which present several features (split, arc) that can disturb the minimization. This model helped us to put constraints on the position and inner slope of the disk. We used this position to improve the constraints already obtained by \cite{Biller2015} on a putative planet carving this inner disk. With both NICMOS/RDI and NICI/ADI images, we are not able to detect any dust inside the cavity within 100 AU of which the presence was inferred by previous authors \citep{Augereau1999_hd141,Thi2014}.

Finally, using a rather aggressive LOCI reduction, we were able to observe fine structures in the disk, previously unreported. Among them, a split in two narrow rings/arcs of the inner disk on the East. Fig~\ref{fig:schema_thick} (right) shows a possible interpretation of this split in the East part of the inner disk. Assuming we are looking at the side of a non-flat optically thick ring and that the central part of this disk (dashed line in Fig~\ref{fig:normalized_contrast}, right) absorbs the light, the radius of the disk is therefore $R =(r_2 + r_1)/2 = 223$ AU, while $H.\sin(i) = (r_2 - r_1) = 21$ AU. We deduced $H = 25$ AU, which corresponds to an opening angle of $h = H/R = 11\%$. However, this explanation is not coherent with previous conclusion from \cite{Thi2014}. These author found a value of $h<10\%$ for the gas, deduced from the analysis in the inner region. More importantly, these authors prove that the disk is optically thin vertically and radially in the optical. This feature is more likely to be an actual split between two coplanar components of the same flat disk. However, the fact that we cannot see it at all position angles tends to prove that the two components of this disk are of different composition. Finally, a small finger is detached from the inner edge, on the South East.

As of early 2015 this disk is now being further observed by several coronagraphic instruments, including VLT/SPHERE \citep{Beuzit2008}, GEMINI/GPI \cite{Macintosh2008} and HST/STIS. These new observations will help to confirm the observed structures in this disk and to detect or put stronger constraints on a putative planet generating these structures. However, the interpretation of these features (strong inner slope, depleted cavity, splitting in the West part of the disk) must also take into account that this disk contains a large amount of gas \citep{Thi2014}. Indeed, \citet{Lyra13} have shown that the presence of gas could also carve sharp eccentric rings in a disk as planets do. Therefore, observations at millimetric wavelengths with ALMA of the gas content will certainly be critical to improve the comprehension of this object. 

\appendix
\section{Appendix}
\label{sec:annex}

\begin{figure}
\begin{center}
 \includegraphics[height=8cm]{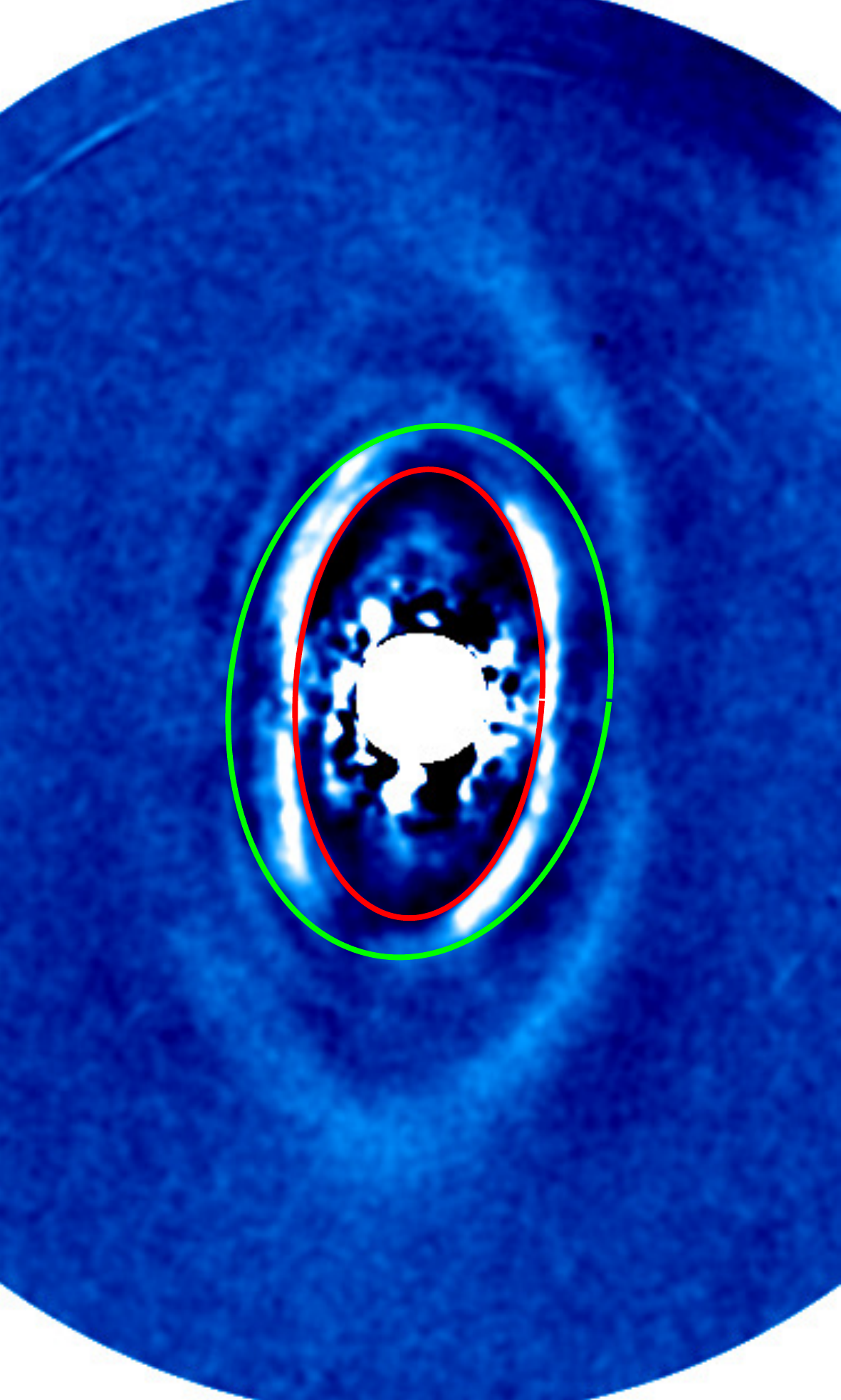}
 \end{center}
\caption[ADI_disk]
{\label{fig:compareBiller} In this figure, we show in red the ellipse found using our argument (already plotted in Fig.~\ref{fig:ellipsefit}, right) and in green the ellipse found using geometrical parameters found in Table 2 of \cite{Biller2015}.}
\end{figure}

In Fig.~\ref{fig:compareBiller}, we show the NICI/cADI image, on which we plotted in red the ellipse found using our argument (the derivative maximum, previously plotted in Fig.~\ref{fig:ellipsefit}, right) and in green the ellipse found using geometrical parameters found in Table 2 of \cite{Biller2015} (inclination = $44.9\degree$, offset X = -2.5 AU, offset Y = 3.1 AU, PA = $351.1\degree$). This green ellipse is not fitting the data, although we are not sure of the reason.

\acknowledgments{J. Mazoyer was supported by the Centre National d'Etudes Spatiales (CNES, Toulouse, France) and Astrium (Toulouse, France) during his PhD. A.~Boccaletti, J.-C.~Augereau, and A.-M.~Lagrange are supported by ANR-14-CE33-0018 and the PNP. The authors would also like to thank the referee for its very interesting comments.}

\bibliographystyle{apj} 
\bibliography{hd141_bib}  

\begin{thebibliography}{}
\expandafter\ifx\csname natexlab\endcsname\relax\def\natexlab#1{#1}\fi

\bibitem[{{Ardila} {et~al.}(2005){Ardila}, {Lubow}, {Golimowski}, {Krist},
  {Clampin}, {Ford}, {Hartig}, {Illingworth}, {Bartko}, {Ben{\'i}tez},
  {Blakeslee}, {Bouwens}, {Bradley}, {Broadhurst}, {Brown}, {Burrows}, {Cheng},
  {Cross}, {Feldman}, {Franx}, {Goto}, {Gronwall}, {Holden}, {Homeier},
  {Infante}, {Kimble}, {Lesser}, {Martel}, {Menanteau}, {Meurer}, {Miley},
  {Postman}, {Sirianni}, {Sparks}, {Tran}, {Tsvetanov}, {White}, {Zheng}, \&
  {Zirm}}]{Ardila2005}
{Ardila}, D.~R., {Lubow}, S.~H., {Golimowski}, D.~A., {et~al.} 2005, \apj, 627,
  986

\bibitem[{{Augereau} {et~al.}(1999{\natexlab{a}}){Augereau}, {Lagrange},
  {Mouillet}, \& {M{\'e}nard}}]{Augereau1999_hd141}
{Augereau}, J.~C., {Lagrange}, A.~M., {Mouillet}, D., \& {M{\'e}nard}, F.
  1999{\natexlab{a}}, \aap, 350, L51

\bibitem[{{Augereau} {et~al.}(1999{\natexlab{b}}){Augereau}, {Lagrange},
  {Mouillet}, {Papaloizou}, \& {Grorod}}]{Augereau1999_hr47_grater}
{Augereau}, J.~C., {Lagrange}, A.~M., {Mouillet}, D., {Papaloizou}, J.~C.~B.,
  \& {Grorod}, P.~A. 1999{\natexlab{b}}, \aap, 348, 557

\bibitem[{{Augereau} \& {Papaloizou}(2004)}]{Augereau_papaloizeau2004}
{Augereau}, J.~C., \& {Papaloizou}, J.~C.~B. 2004, \aap, 414, 1153

\bibitem[{{Baraffe} {et~al.}(2002){Baraffe}, {Chabrier}, {Allard}, \&
  {Hauschildt}}]{Baraffe2002}
{Baraffe}, I., {Chabrier}, G., {Allard}, F., \& {Hauschildt}, P.~H. 2002, \aap,
  382, 563

\bibitem[{{Baraffe} {et~al.}(2003){Baraffe}, {Chabrier}, {Barman}, {Allard}, \&
  {Hauschildt}}]{Baraffe2003}
{Baraffe}, I., {Chabrier}, G., {Barman}, T.~S., {Allard}, F., \& {Hauschildt},
  P.~H. 2003, \aap, 402, 701

\bibitem[{{Beuzit} {et~al.}(2008){Beuzit}, {Feldt}, {Dohlen}, {Mouillet},
  {Puget}, {Wildi}, {Abe}, {Antichi}, {Baruffolo}, {Baudoz}, {Boccaletti},
  {Carbillet}, {Charton}, {Claudi}, {Downing}, {Fabron}, {Feautrier},
  {Fedrigo}, {Fusco}, {Gach}, {Gratton}, {Henning}, {Hubin}, {Joos}, {Kasper},
  {Langlois}, {Lenzen}, {Moutou}, {Pavlov}, {Petit}, {Pragt}, {Rabou}, {Rigal},
  {Roelfsema}, {Rousset}, {Saisse}, {Schmid}, {Stadler}, {Thalmann}, {Turatto},
  {Udry}, {Vakili}, \& {Waters}}]{Beuzit2008}
{Beuzit}, J.-L., {Feldt}, M., {Dohlen}, K., {et~al.} 2008, in Society of
  Photo-Optical Instrumentation Engineers (SPIE) Conference Series, Vol. 7014,
  Society of Photo-Optical Instrumentation Engineers (SPIE) Conference Series

\bibitem[{{Biller} {et~al.}(2015){Biller}, {Liu}, {Rice}, {Wahhaj}, {Nielsen},
  {Hayward}, {Kuchner}, {Close}, {Chun}, {Ftaclas}, \& {Toomey}}]{Biller2015}
{Biller}, B.~A., {Liu}, M.~C., {Rice}, K., {et~al.} 2015, \mnras, 450, 4446

\bibitem[{{Boccaletti} {et~al.}(2003){Boccaletti}, {Augereau}, {Marchis}, \&
  {Hahn}}]{Boccaletti2003}
{Boccaletti}, A., {Augereau}, J.-C., {Marchis}, F., \& {Hahn}, J. 2003, \apj,
  585, 494

\bibitem[{Boccaletti {et~al.}(2013{\natexlab{a}})Boccaletti, Lagrange,
  Bonnefoy, Galicher, \& Chauvin}]{Boccaletti2013a}
Boccaletti, A., Lagrange, A.~M., Bonnefoy, M., Galicher, R., \& Chauvin, G.
  2013{\natexlab{a}}, Astronomy {\&} Astrophysics, 551, L14

\bibitem[{Boccaletti {et~al.}(2013{\natexlab{b}})Boccaletti, Pantin, Lagrange,
  Augereau, Meheut, \& Quanz}]{Boccaletti2013b}
Boccaletti, A., Pantin, E., Lagrange, A.~M., {et~al.} 2013{\natexlab{b}},
  Astronomy {\&} Astrophysics, 560, A20

\bibitem[{{Casassus} {et~al.}(2013){Casassus}, {van der Plas}, {M}, {Dent},
  {Fomalont}, {Hagelberg}, {Hales}, {Jord{\'a}n}, {Mawet}, {M{\'e}nard},
  {Wootten}, {Wilner}, {Hughes}, {Schreiber}, {Girard}, {Ercolano}, {Canovas},
  {Rom{\'a}n}, \& {Salinas}}]{Casassus2013}
{Casassus}, S., {van der Plas}, G., {M}, S.~P., {et~al.} 2013, \nat, 493, 191

\bibitem[{{Choquet} {et~al.}(2014){Choquet}, {Pueyo}, {Hagan}, {Gofas-Salas},
  {Rajan}, {Chen}, {Perrin}, {Debes}, {Golimowski}, {Hines}, {N'Diaye},
  {Schneider}, {Mawet}, {Marois}, \& {Soummer}}]{Choquet2014}
{Choquet}, {\'E}., {Pueyo}, L., {Hagan}, J.~B., {et~al.} 2014, in Society of
  Photo-Optical Instrumentation Engineers (SPIE) Conference Series, Vol. 9143,
  Society of Photo-Optical Instrumentation Engineers (SPIE) Conference Series,
  57

\bibitem[{{Clampin} {et~al.}(2003){Clampin}, {Krist}, {Ardila}, {Golimowski},
  {Hartig}, {Ford}, {Illingworth}, {Bartko}, {Ben{\'{\i}}tez}, {Blakeslee},
  {Bouwens}, {Broadhurst}, {Brown}, {Burrows}, {Cheng}, {Cross}, {Feldman},
  {Franx}, {Gronwall}, {Infante}, {Kimble}, {Lesser}, {Martel}, {Menanteau},
  {Meurer}, {Miley}, {Postman}, {Rosati}, {Sirianni}, {Sparks}, {Tran},
  {Tsvetanov}, {White}, \& {Zheng}}]{Clampin2003}
{Clampin}, M., {Krist}, J.~E., {Ardila}, D.~R., {et~al.} 2003, \aj, 126, 385

\bibitem[{Currie {et~al.}(2012)Currie, Rodigas, Debes, Plavchan, Kuchner,
  Jang-Condell, Wilner, Andrews, Kraus, Dahm, \& Robitaille}]{Currie2012}
Currie, T., Rodigas, T.~J., Debes, J., {et~al.} 2012, The Astrophysical
  Journal, 757, 28

\bibitem[{{Fisher} {et~al.}(2000){Fisher}, {Telesco}, {Pi{\~n}a}, {Knacke}, \&
  {Wyatt}}]{Fisher2000}
{Fisher}, R.~S., {Telesco}, C.~M., {Pi{\~n}a}, R.~K., {Knacke}, R.~F., \&
  {Wyatt}, M.~C. 2000, \apjl, 532, L141

\bibitem[{{Goto} {et~al.}(2006){Goto}, {Usuda}, {Dullemond}, {Henning}, {Linz},
  {Stecklum}, \& {Suto}}]{Goto2006}
{Goto}, M., {Usuda}, T., {Dullemond}, C.~P., {et~al.} 2006, \apj, 652, 758

\bibitem[{{Lafreni{\`e}re} {et~al.}(2007){Lafreni{\`e}re}, {Marois}, {Doyon},
  {Nadeau}, \& {Artigau}}]{Lafreniere2007}
{Lafreni{\`e}re}, D., {Marois}, C., {Doyon}, R., {Nadeau}, D., \& {Artigau},
  {\'E}. 2007, \apj, 660, 770

\bibitem[{Lagrange {et~al.}(2009)Lagrange, Gratadour, Chauvin, Fusco,
  Ehrenreich, Mouillet, Rousset, Rouan, Allard, Gendron, Charton, Mugnier,
  Rabou, Montri, \& Lacombe}]{Lagrange2009}
Lagrange, A.~M., Gratadour, D., Chauvin, G., {et~al.} 2009, Astronomy {\&}
  Astrophysics, 493, L21

\bibitem[{{Lagrange} {et~al.}(2012){Lagrange}, {Boccaletti}, {Milli},
  {Chauvin}, {Bonnefoy}, {Mouillet}, {Augereau}, {Girard}, {Lacour}, \&
  {Apai}}]{Lagrange2012}
{Lagrange}, A.-M., {Boccaletti}, A., {Milli}, J., {et~al.} 2012, \aap, 542, A40

\bibitem[{{Lebreton} {et~al.}(2013){Lebreton}, {van Lieshout}, {Augereau},
  {Absil}, {Mennesson}, {Kama}, {Dominik}, {Bonsor}, {Vandeportal}, {Beust},
  {Defr{\`e}re}, {Ertel}, {Faramaz}, {Hinz}, {Kral}, {Lagrange}, {Liu}, \&
  {Th{\'e}bault}}]{Lebreton2013}
{Lebreton}, J., {van Lieshout}, R., {Augereau}, J.-C., {et~al.} 2013, \aap,
  555, A146

\bibitem[{{Lyra} \& {Kuchner}(2013)}]{Lyra13}
{Lyra}, W., \& {Kuchner}, M. 2013, \nat, 499, 184

\bibitem[{{Macintosh} {et~al.}(2008){Macintosh}, {Graham}, {Palmer}, {Doyon},
  {Dunn}, {Gavel}, {Larkin}, {Oppenheimer}, {Saddlemyer}, {Sivaramakrishnan},
  {Wallace}, {Bauman}, {Erickson}, {Marois}, {Poyneer}, \&
  {Soummer}}]{Macintosh2008}
{Macintosh}, B.~A., {Graham}, J.~R., {Palmer}, D.~W., {et~al.} 2008, in Society
  of Photo-Optical Instrumentation Engineers (SPIE) Conference Series, Vol.
  7015, Society of Photo-Optical Instrumentation Engineers (SPIE) Conference
  Series

\bibitem[{{Marois} {et~al.}(2006){Marois}, {Lafreni{\`e}re}, {Doyon},
  {Macintosh}, \& {Nadeau}}]{Marois2006}
{Marois}, C., {Lafreni{\`e}re}, D., {Doyon}, R., {Macintosh}, B., \& {Nadeau},
  D. 2006, \apj, 641, 556

\bibitem[{Marois {et~al.}(2008)Marois, Lafreni{\`e}re, Macintosh, \&
  Doyon}]{Marois2008}
Marois, C., Lafreni{\`e}re, D., Macintosh, B., \& Doyon, R. 2008, The
  Astrophysical Journal, 673, 647

\bibitem[{{Mazoyer} {et~al.}(2014){Mazoyer}, {Boccaletti}, {Augereau},
  {Lagrange}, {Galicher}, \& {Baudoz}}]{Mazoyer2014}
{Mazoyer}, J., {Boccaletti}, A., {Augereau}, J.-C., {et~al.} 2014, \aap, 569,
  A29

\bibitem[{Milli {et~al.}(2012)Milli, Mouillet, Lagrange, Boccaletti, Mawet,
  Chauvin, \& Bonnefoy}]{Milli2012}
Milli, J., Mouillet, D., Lagrange, A.~M., {et~al.} 2012, Astronomy {\&}
  Astrophysics, 545, A111

\bibitem[{{Moerchen} {et~al.}(2010){Moerchen}, {Telesco}, \&
  {Packham}}]{Moerchen10}
{Moerchen}, M.~M., {Telesco}, C.~M., \& {Packham}, C. 2010, \apj, 723, 1418

\bibitem[{{Mouillet} {et~al.}(2001){Mouillet}, {Lagrange}, {Augereau}, \&
  {M{\'e}nard}}]{Mouillet2001}
{Mouillet}, D., {Lagrange}, A.~M., {Augereau}, J.~C., \& {M{\'e}nard}, F. 2001,
  \aap, 372, L61

\bibitem[{{Mulders} {et~al.}(2013){Mulders}, {Min}, {Dominik}, {Debes}, \&
  {Schneider}}]{Mulders2013}
{Mulders}, G.~D., {Min}, M., {Dominik}, C., {Debes}, J.~H., \& {Schneider}, G.
  2013, \aap, 549, A112

\bibitem[{{Ozernoy} {et~al.}(2000){Ozernoy}, {Gorkavyi}, {Mather}, \&
  {Taidakova}}]{Ozernoy2000}
{Ozernoy}, L.~M., {Gorkavyi}, N.~N., {Mather}, J.~C., \& {Taidakova}, T.~A.
  2000, \apjl, 537, L147

\bibitem[{Racine {et~al.}(1999)Racine, Walker, Nadeau, Doyon, \&
  Marois}]{Racine1999}
Racine, R., Walker, G. A.~H., Nadeau, D., Doyon, R., \& Marois, C. 1999, The
  Publications of the Astronomical Society of the Pacific, 111, 587

\bibitem[{{Rameau} {et~al.}(2013){Rameau}, {Chauvin}, {Lagrange}, {Meshkat},
  {Boccaletti}, {Quanz}, {Currie}, {Mawet}, {Girard}, {Bonnefoy}, \&
  {Kenworthy}}]{Rameau2013}
{Rameau}, J., {Chauvin}, G., {Lagrange}, A.-M., {et~al.} 2013, \apjl, 779, L26

\bibitem[{{Reche} {et~al.}(2009){Reche}, {Beust}, \& {Augereau}}]{Reche2009}
{Reche}, R., {Beust}, H., \& {Augereau}, J.-C. 2009, \aap, 493, 661

\bibitem[{Rodigas {et~al.}(2012)Rodigas, Hinz, Leisenring, Vaitheeswaran,
  Skemer, Skrutskie, Su, Bailey, Schneider, Close, Mannucci, Esposito,
  Arcidiacono, Pinna, Argomedo, Agapito, Apai, Bono, Boutsia, Briguglio, Brusa,
  Busoni, Cresci, Currie, Desidera, Eisner, Falomo, Fini, Follette, Fontana,
  Garnavich, Gratton, Green, Guerra, Hill, Hoffmann, Jones, Krejny, Kulesa,
  Males, Masciadri, Mesa, McCarthy, Meyer, Miller, Nelson, Puglisi,
  Quiros-Pacheco, Riccardi, Sani, Stefanini, Testa, Wilson, Woodward, \&
  Xompero}]{Rodigas2012}
Rodigas, T.~J., Hinz, P.~M., Leisenring, J., {et~al.} 2012, The Astrophysical
  Journal, 752, 57

\bibitem[{{Schneider} {et~al.}(2010){Schneider}, {Silverstone}, {Stobie},
  {Rhee}, \& {Hines}}]{Schneider2010}
{Schneider}, G., {Silverstone}, M.~D., {Stobie}, E., {Rhee}, J.~H., \& {Hines},
  D.~C. 2010, in Hubble after SM4. Preparing JWST, 15

\bibitem[{{Soummer} {et~al.}(2011){Soummer}, {Brendan Hagan}, {Pueyo},
  {Thormann}, {Rajan}, \& {Marois}}]{Soummer2011}
{Soummer}, R., {Brendan Hagan}, J., {Pueyo}, L., {et~al.} 2011, \apj, 741, 55

\bibitem[{Soummer {et~al.}(2012)Soummer, Pueyo, \& Larkin}]{Soummer2012}
Soummer, R., Pueyo, L., \& Larkin, J. 2012, The Astrophysical Journal Letters,
  755, L28

\bibitem[{{Soummer} {et~al.}(2014){Soummer}, {Perrin}, {Pueyo}, {Choquet},
  {Chen}, {Golimowski}, {Brendan Hagan}, {Mittal}, {Moerchen}, {N'Diaye},
  {Rajan}, {Wolff}, {Debes}, {Hines}, \& {Schneider}}]{Soummer2014}
{Soummer}, R., {Perrin}, M.~D., {Pueyo}, L., {et~al.} 2014, \apjl, 786, L23

\bibitem[{Thalmann {et~al.}(2010)Thalmann, Grady, Goto, Wisniewski, Janson,
  Henning, Fukagawa, Honda, Mulders, Min, Moro-Mart{\'\i}n, McElwain, Hodapp,
  Carson, Abe, Brandner, Egner, Feldt, Fukue, Golota, Guyon, Hashimoto, Hayano,
  Hayashi, Hayashi, Ishii, Kandori, Knapp, Kudo, Kusakabe, Kuzuhara, Matsuo,
  Miyama, Morino, Nishimura, Pyo, Serabyn, Shibai, Suto, Suzuki, Takami,
  Takato, Terada, Tomono, Turner, Watanabe, Yamada, Takami, Usuda, \&
  Tamura}]{Thalmann2010}
Thalmann, C., Grady, C.~A., Goto, M., {et~al.} 2010, The Astrophysical Journal
  Letters, 718, L87

\bibitem[{{Thi} {et~al.}(2014){Thi}, {Pinte}, {Pantin}, {Augereau}, {Meeus},
  {M{\'e}nard}, {Martin-Za{\"i}di}, {Woitke}, {Riviere-Marichalar}, {Kamp},
  {Carmona}, {Sandell}, {Eiroa}, {Dent}, {Montesinos}, {Aresu}, {Meijerink},
  {Spaans}, {White}, {Ardila}, {Lebreton}, {Mendigut{\'{\i}}a}, \&
  {Brittain}}]{Thi2014}
{Thi}, W.-F., {Pinte}, C., {Pantin}, E., {et~al.} 2014, \aap, 561, A50

\bibitem[{Toomey \& Ftaclas(2003)}]{Toomey2003}
Toomey, D.~W., \& Ftaclas, C. 2003, Instrument Design and Performance for
  Optical/Infrared Ground-based Telescopes. Edited by Iye, 4841, 889

\bibitem[{{van Leeuwen}(2007)}]{vanLeeuwen2007}
{van Leeuwen}, F. 2007, \aap, 474, 653

\bibitem[{Wahhaj {et~al.}(2011)Wahhaj, Liu, Biller, Clarke, Nielsen, Close,
  Hayward, Mamajek, Cushing, Dupuy, Tecza, Thatte, Chun, Ftaclas, Hartung,
  Reid, Shkolnik, Alencar, Artymowicz, Boss, de~Gouveia Dal~Pino,
  Gregorio-Hetem, Ida, Kuchner, Lin, \& Toomey}]{Wahhaj2011}
Wahhaj, Z., Liu, M.~C., Biller, B.~A., {et~al.} 2011, The Astrophysical
  Journal, 729, 139

\bibitem[{{Wahl} {et~al.}(2013){Wahl}, {Metchev}, {Patel}, {Serabyn}, \&
  {PALM-3000 Adaptive Optics Team}}]{Wahl2013}
{Wahl}, M., {Metchev}, S.~A., {Patel}, R., {Serabyn}, G., \& {PALM-3000
  Adaptive Optics Team}. 2013, in American Astronomical Society Meeting
  Abstracts, Vol. 221, American Astronomical Society Meeting Abstracts \#221,
  144.22

\bibitem[{{Weinberger} {et~al.}(1999){Weinberger}, {Becklin}, {Schneider},
  {Smith}, {Lowrance}, {Silverstone}, {Zuckerman}, \&
  {Terrile}}]{Weinberger1999}
{Weinberger}, A.~J., {Becklin}, E.~E., {Schneider}, G., {et~al.} 1999, \apjl,
  525, L53

\bibitem[{{Weinberger} {et~al.}(2000){Weinberger}, {Rich}, {Becklin},
  {Zuckerman}, \& {Matthews}}]{Weinberger2000}
{Weinberger}, A.~J., {Rich}, R.~M., {Becklin}, E.~E., {Zuckerman}, B., \&
  {Matthews}, K. 2000, \apj, 544, 937

\bibitem[{{Wisdom}(1980)}]{Wisdom1980}
{Wisdom}, J. 1980, \aj, 85, 1122

\bibitem[{{Wyatt}(2005)}]{Wyatt2005}
{Wyatt}, M.~C. 2005, \aap, 440, 937

\end{thebibliography}

\end{document}